# High-speed surface-property recognition by 140-GHz frequency

Jiacheng Liu *Student Member, IEEE,* Da Li, Guohao Liu, Yige Qiao *Student Member, IEEE,* Menghan Wei, Chengyu Zhang, Fei Song, Jianjun Ma *Member, IEEE*

*Abstract*—In the field of integrated sensing and communication, there's a growing need for advanced environmental perception. The terahertz (THz) frequency band, significant for ultra-high-speed data connections, shows promise in environmental sensing, particularly in detecting surface textures crucial for autonomous systems' decision-making. However, traditional numerical methods for parameter estimation in these environments struggle with accuracy, speed, and stability, especially in high-speed scenarios like vehicle-to-everything communications. This study introduces a deep learning approach for identifying surface roughness using a 140-GHz setup tailored for high-speed conditions. A high-speed data acquisition system was developed to mimic real-world scenarios, and a diverse set of rough surface samples was collected for realistic high-speed datasets to train the models. The model was trained and validated in three challenging scenarios: *random occlusions*, *sparse data*, and *narrow-angle observations*. The results demonstrate the method's effectiveness in high-speed conditions, suggesting terahertz frequencies' potential in future sensing and communication applications

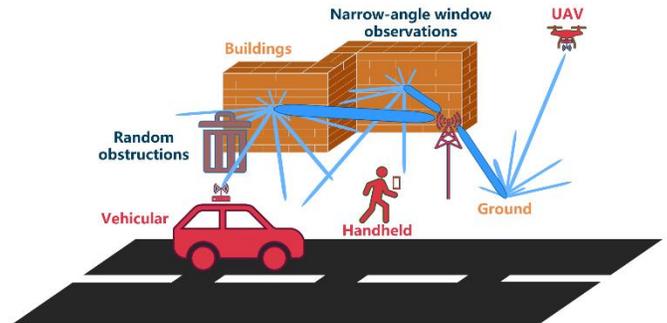

**Fig. 1.** A schematic of high-speed scenarios, including handheld, vehicle mounted, and Unmanned Aerial Vehicle (UAV) scenarios.

## I. INTRODUCTION

As the integration of sensing and communication technologies rapidly advances, the terahertz (THz) frequency band has emerged as a pivotal frontier for both domains, presents opportunities in achieving high data rate communication and precise environmental sensing[1], [2]. These features promise a significant advancement in capabilities for 6G integrated sensing and communication (ISAC) technology [3], autonomous vehicles [4] and vehicular communication networks (VCN) [5] In such integrated systems, the role of environmental sensing becomes important increasingly. They require swift and precise perception of the environment to adapt in real-time environmental variations, and enhance communication and/or control system stability [6], [7]. In these scenarios, accurate recognition of surface roughness emerges as a critical environmental parameter. The scattering characteristics of wireless channels, for an example, are profoundly influenced by the texture of surrounding surfaces[8], and the control system of autonomous vehicles must adapt to the roughness of the terrain [9]. Terahertz frequencies are inherently sensitive to surface granularity, offering a potent advantage in addressing these challenges [10].

Terahertz frequencies can decode intricate details of surrounding surfaces, yielding distinct signatures within their scattering spectra [11], [12]. This makes them invaluable for non-invasive surface texture assessments. While diverse strategies, such as Kirchhoff theory [13], Rayleigh roughness theory [14], time-of-flight method [15], etc., have been developed for terahertz surface texture modeling and recognition, relying on numerical scattering models [16], [17]. These methods require complicated modeling trajectories and are anchored in approximate and idealized postulates, which often fail to accurately capture real-world complexities. Other techniques harness the richness of terahertz imaging features, predominantly integrating deep learning frameworks [9], [18], [19]. While these approaches are adept in low-speed scenarios, they are not well-suited for high-speed conditions, where numerical computations become impractical and radar imagery becomes prohibitively resource-intensive.

In high-speed scenarios, static equipment positioning is unfeasible, and sampling must transpire dynamically. This introduces tailing effect, which means the data captured at a specific time and angle includes traces from previous conditions. The magnitude of tailing effect increases with speed, leading to directional spectral biases and attenuation of spectral details. Besides, high-speed sampling introduces

This work was supported in part by the National Natural Science Foundation of China (62071046), the Science and Technology Innovation Program of Beijing Institute of Technology (2022CX01023), the Graduate Innovative Practice Project of Tangshan Research Institute, BIT (TSDZXX202201) and the Talent Support Program of Beijing Institute of Technology "Special Young Scholars" (3050011182153). (Corresponding: Jianjun Ma)

Jiacheng Liu, Da Li, Yige Qiao, Menghan Wei are with Beijing Institute of Technology, Beijing, 100081 China (e-mail: jia.cheng.liu, 3220221512, 3220210640, 3120210675@bit.edu.cn).

Guohao Liu, Jianjun Ma are with Beijing Institute of Technology, Beijing, 100081 China, and Tangshan Research Institute, BIT, Tangshan, Hebei, 063099 China (e-mail: 3220221571, jianjun_ma@bit.edu.cn).

Chengyu Zhang is with North Automatic Control Technology Institute, Taiyuan, Shanxi, 030000 China (e-mail: zhangchengyu010@163.com).

Fei Song is with Beijing Jiaotong University, Beijing, 100044 China (e-mail: fsong@bjtu.edu.cn).

additional challenges, such as *random occlusions*, *high-speed sparse data*, and *narrow-angle window observations*, which render conventional methodologies inadequate [20]. To overcome these challenges, we propose a deep learning-driven paradigm for surface texture recognition in high-speed scenarios, which has already been demonstrated as its prowess in pragmatic terahertz applications such as imaging [21], synthetic aperture radar (SAR) [22], and hyperspectral modalities [23].

However, leveraging deep learning for surface texture recognition in high-speed conditions presents several limitations in: (1) Data source: Preliminary trials revealed that data generated from simulation models often bears an idealized guise, rendering it amenable for deep learning assimilation, yet potentially misaligned with real-world dynamics [24]. Authentic data acquisition becomes non-negotiable. (2) Data diversity: To ensure the model's extrapolative capabilities posttraining on a constrained roughness dataset, a rich and varied training dataset is mandatory. (3) Efficiency of model training and prediction. The purpose lies in adeptly training the model on a comparatively concise dataset, ensuring its robustness in intricate scenarios. To overcome these, a high-speed data recording platform that emulating real-world dynamics was designed. We fabricated several kinds of rough surface samples, characterized them by high-fidelity 3D scanning, and acquired lots of high-speed scattering data. To strengthen data heterogeneity, we focused on localized sample regions and further enriched them with diverse data processing and augmentation strategies to sculpt the dataset.

The following of this article is laid out as: Section II gives a thorough overview of our data collection platform, which includes how we set up our experiments and the system we used to control them. In Section III shows a detailed description on how we created and assessed the rough surfaces that we used as samples. Section IV explains how we recorded the data needed to train our algorithms. Section V describes how we designed the algorithm that our network employed. Section VI discusses the experiments we conducted using deep learning and delves into what the results mean. Finally, Section VII concludes the entire work. We think this work presents a significant advancement in the field of surface texture recognition by terahertz frequencies, enabling robust and reliable operation in high-speed scenarios. We also think it opens up new possibilities for a wide range of applications, including autonomous driving, environmental monitoring, and industrial inspection.

## II. DATA SAMPLING PLATFORMS

To meet the demands of real-world applications, we fabricated a high-speed sampling platform, depicted in Fig. 2(a). This platform is designed to fulfill the dual criteria of rapid rotation and high precision, achieved through a fully automated programming-driven sampling system. Its core is a high-precision electrical control rotational stage, powered by a 57-step stepper motor, and governed by a worm gear structure. The motor, with a step angle of 1.8° and a mechanical transmission ratio of 1:180, confirms the rotational stage with a minimum rotation angle of 0.01°. This makes its precision and alignment under the requirements in measurement.

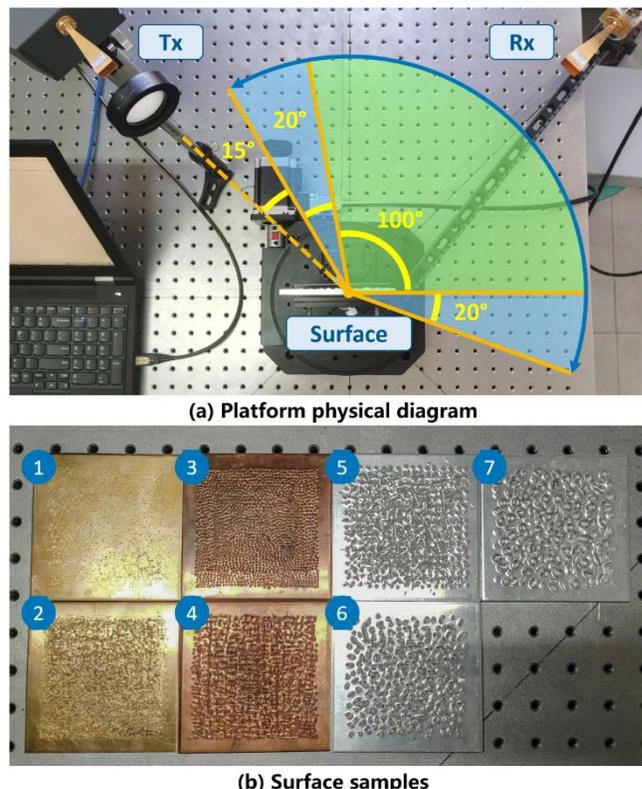

Fig. 2. **(a)** The image of the data sampling platforms, including an electric rotational stage that drives Rx in a circular motion. The blue arrows indicate the range of motion for Rx, which includes variable speed regions of 20° each in the front and rear, as well as a uniform speed region of 100° in the middle. **(b)** Surface samples created using a combination of materials, tools, and shapes.

The experimental setup involved affixing the receiver (Rx) onto the rotational stage via an extended rail, placing the rough surface sample at the rotational stage's epicenter, and situating the transmitter (Tx) at a 45° incident angle. The spatial arrangement saw the transmitter and receiver positioned 36cm and 38cm away from the sample's geometric center, respectively. The transmission assembly comprises a signal generator (Ceyear 1465D), a frequency multiplier module (Ceyear 82406B), and a horn antenna. The signal generator (Ceyear 1465), has the capability to generate signals up to 20 GHz, which are then up-converted to 110-170 GHz range by the Ceyear 82406D. The receiver employs an identical horn antenna to capture signals, and forwards them to a power sensor (Ceyear 71718) for signal detection, with a single sampling time of 50ms. To enhance focusing, a lens with a 10cm focal length was strategically positioned in front of Tx, aligning with the focal point. Throughout the sampling process, while Tx and the surface sample remained stationary, Rx embarked on a circular trajectory around the sample, instigating variations in reflection angles. Both Tx and Rx are fixed within a horizontal plane, which is not fully representative of practical situations where an elevation angle may be often present [24], [25]. However, for environmental sensing in the real world, we think that a near-zero elevation angle can be sufficient in most cases.

The electrical control rotational stage was exercised through a serial port protocol, ensuring continuous motion data monitoring. Concurrently, LAN and SCPI protocols were deployed for the terahertz signal generator and power meter management. The amalgamation of these components was seamlessly orchestrated using Python3.9, which facilitated interface calls, ensured time synchronization, governed sampling control, and oversaw data storage operations. This sophisticated platform guarantees an automatic and highly efficient data acquisition process with a communication latency less than 10ms

## III. SURFACE SAMPLE PREPARATION

Data is paramount in the realm of neural networks. Aiming for effective performance on real-world surfaces with unknown roughness, it is important to improve the generalization capabilities of limited set of surface samples with different levels of roughness. This requires our samples to embody typical characteristic properties. Due to the difficulties in obtaining enough data for our investigation, we chose metal surfaces for our primary samples, due to their excellent reflective properties and a diverse range of hardnesses. We applied different processes to forge surfaces with different levels of roughness, as shown in Fig. 2(b).

To create smoother surfaces, resembling scratches and minor textural variations on office desks or brushed finishes, we used small-diameter rectangular and cylindrical chisels on harder brass materials to create two distinct surfaces. Surface1 features scratches and minor abrasions, while Surface-2 predominantly features scratches with shallow grooves. To create rougher surfaces with inherent natural textures, such as wood grain and genuine leather patterns, we used conical and medium-diameter rectangular chisels on materials of moderate hardness, specifically copper. Surface-3 is composed of dense conical pits, emulating certain grid-like textures. Surface-4, derived from Surface-3, incorporates deeper grooves to mimic deep leather textures. Given that rougher surfaces often exhibit more pronounced relief structures, we used larger-diameter conical and rectangular chisels on softer aluminum sheets to produce three additional surfaces. Surface-5 is an enhanced version of Surface-4. Surfaces-6 and 7, in contrast to the previous methods of vertical carving, employ lateral force application to create substantial depressions and peaks in the structure. All these surface samples own a large size, measuring 100mm×100mm. Within this area, the central region of 80mm×80mm exhibited the roughness characteristics. We used a grid counting method to ensure maximum uniformity in the local roughness within this region. Such a large surface area endows the dataset with a richer diversity of data, featuring similar numerical roughness but varied structures.

TABLE I
ORIGINAL DATA

| Idx. | Rotational Speed (°/s) | Averages # | Query Interval (s) | Cumulative length | Counts |
|---|---|---|---|---|---|
| 1 | 20 | 4 | 0.1 | 55 | 504 |
| 2 | 20 | 4 | 0.05 | 85 | 504 |
| 3 | 20 | 1 | 0.05 | 85 | 504 |
| 4 | 10 | 4 | 0.1 | 110 | 504 |
| 5 | 10 | 1 | 0.05 | 170 | 504 |

For accurate characterization of the surface samples, we conducted a comprehensive 3-dimensional (3D) laser scanning (SCANTECH IREAL2E) using high-precision techniques to reconstruct all the surfaces. We utilized an industrial-grade handheld 3D laser scanner with a scanning precision of 0.02mm and a volumetric accuracy of 0.03mm/m. Each metal surface underwent scanning to produce at least two million-point clouds, effectively representing the roughness of various local areas on the surface. We used the standard deviation of the surface height variation (RMS) as a quantitative measure of surface roughness [26]. The ensuing results, detailed in Fig. 6 of the Appendix part, reveal that our samples of rough surfaces manifest several characteristic traits: (1) Minimal overlap of roughness between different surfaces, which is essential for subsequent classification tasks, yielding classification results that can be uniquely mapped back to a specific range of roughness. (2) Variability in roughness across different areas within a single surface. This enriches the dataset and enhances the model's ability to generalize. (3) Existence of regions with comparable levels of roughness, yet differing in shapes and quantities: This further contributes to the overall data richness.

## IV. DATASET ACQUISITION AND COMPOSITION

The data acquisition was conducted by employing the platform and surfaces as in Fig. 2. In low-speed scenarios, scatter sampling is typically performed at one angle per pause (of the rotational stage), ensuring precision and stability. However, high-speed scenarios require continuous sampling, which introduces trailing effect. To emulate real-world highspeed conditions, we continuously sampled while the Rx rotated without pausing. We set the sampling speed to match the rotational stage's safety speed limit, while maintaining a minimum angular distance of 15° between the Rx and Tx. We also included acceleration and deceleration zones (~20°) at the beginning and end of the motion.

In order to improve the robustness and generalization ability of our model, we need a diverse dataset from a limited number of rough surface samples. We achieved this by methodically moving the incident beam across the surface to gather sufficient data for model training. Considering the illumination area (by the channel beam) with a diameter of 3cm on the surface, we moved the beam in both horizontal and vertical dimensions in 5mm increments. With each movement, we collected a set of scattering data, referred as a sampling sequence. Additionally, to mitigate the peak displacement resulting from the tailing effect, we rotated our receiver (RX) in both clockwise and counterclockwise directions at each sampling location. By doing so, we acquired 72 sampling sequences for each surface. In total, this process yielded 504 unique sampling sequences across seven different surfaces.

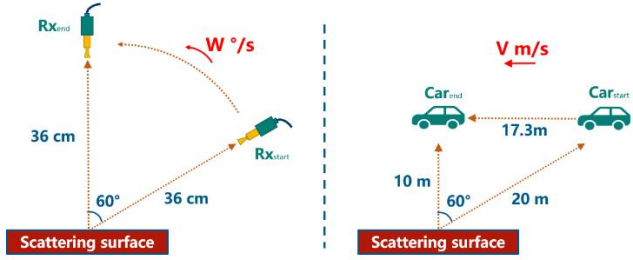

**Fig. 3.** Illustration of the **(a)** high-speed platform scanning scenario and **(b)** vehicle scanning scenario. The vehicle can be replaced with handheld walking or aerial drone flight as the speed changes. The data in the scenario represents only approximate proportions.

To demonstrate the effect of rotation speed of the stage and sampling parameters (average number and query interval), we conducted a comprehensive analysis using a range of speed and parameter settings, as summarized in Table I. Rotation speed refers to the speed at which the motorized rotation stage moves during the uniform motion phase. Average number represents the numerical configurations for the built-in averaging feature of the sampling apparatus. It has been confirmed that its variation within a numerical range has a negligible impact on the final classification F1 score. Query frequency indicates how often the Python main control program retrieves data from the sampling device. It can be seen that rotation speed and query interval (frequency) can affect the cumulative length of the acquired sequences, including data from the acceleration and uniform motion zones near the Tx side.

## V. Scenario and Network Analysis

We described the actual vehicular scenario that our experimental setup aims to emulate, as shown in Fig. 3. For the condition of a vehicle-mounted terahertz sensor traveling through an urban environment, actively scanning building facades. The vertical distance between the target wall surface and the vehicle's trajectory is assumed to be 10m. To avoid the challenges of excessive distances, we focused on a detection range where the incident angle does not exceed 60°. This corresponds to a horizontal span of 17.3m. With a turntable rotating at an angular speed of 10°/s, it takes 6s to traverse a range of 60°. This corresponds to a vehicular speed of approximately 10km/h, which is similar to the pace of a pedestrian. For a rotational speed of 20°/s, the emulated vehicular speed increases to around 20km/h, which is similar to the speed of a vehicle in low-speed cruising.

Using the raw dataset we collected, we have performed a classification task to distinguish between different rough surfaces. The model takes scattering sequence data (X) from a specific position on a surface as input. Each sequence comprises scattering data captured at different angles, represented as X = [$x_0$, $x_1$, ..., $x_N$], where N is the number of sampling points in one sequence. Due to the complexities of real-world high-speed scenarios, it is impractical to assume that the data points adhere to a specific angle-related sequence. Therefore, it is necessary to combine the numerical representations of the sampling points and scattering angles, resulting in the representation $x_t$ = [$\theta_t$, $V_t$]. The model's output is a probability distribution over different surface categories, expressed as Y = [$y_0$, $y_1$, ..., $y_M$], where M is the total number of surface categories.

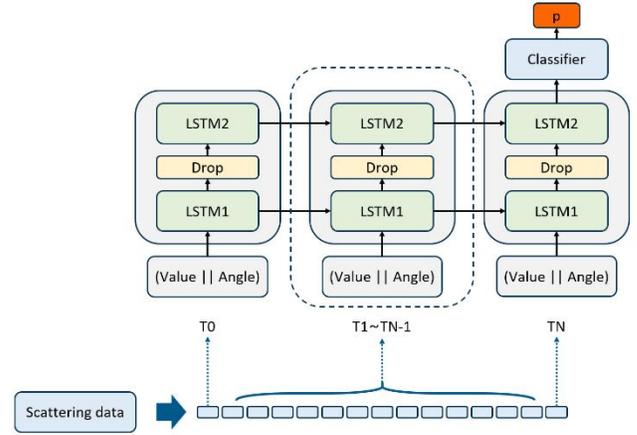

**Fig. 4.** Network architecture diagram.

To enhance the execution of sequence tasks, addressing the inherent variability in sequence lengths is crucial, we use a highly stable LSTM model [27], which is a type of recurrent neural network (RNN). LSTM models have unique gated structures that effectively filter and process information, mitigating redundancy in memory storage. This architecture is particularly well-suited for tasks involving lengthy sequences, and it also solves the vanishing gradient problem that is prevalent in traditional RNNs. The LSTM architecture has three key gate mechanisms: the input gate ($i_t$), the output gate ($i_o$), and the forget gate ($i_f$), as well as a memory cell ($m_t$). At each time step, the model updates the states of these gates based on the current input and output from the previous time step. This process determines which pieces of information should be retained, discarded, or produced as output. The variable $g_t$ represents the input information.

$$i_t = \sigma(W_i x_t + U_i h_{t-1} + b_i) \#(1)$$
$$f_t = \sigma(W_f x_t + U_f h_{t-1} + b_f) \#(2)$$
$$o_t = \sigma(W_o x_t + U_o h_{t-1} + b_o) \#(3)$$
$$g_t = \phi(W_g x_t + U_g h_{t-1} + b_t) \#(4)$$

with σ(.) representing a logistic sigmoid function and ϕ(.) for a hyperbolic tangent function (tanh). Next, the memory cell, mt is updated based on the input and forget gates. The learnable parameters in these expressions are represented by W, U, and b.

$$m_t = f_t \odot m_{t-1} + i_t \odot g_t \#(5)$$

Here ⊙ denotes element-wise product operation. Finally,

the model determines what to output based on the memory and output gates.

$$h_t = o_t \odot \phi(m_t) \quad \#(6)$$

We implemented a multi-layered LSTM network with Dropout layers [28] to mitigate overfitting, the overall architecture of our model is depicted in the Fig. 4. The LSTM layers are followed by a Multi-Layer Perceptron (MLP) classification head, which performs the final classification task. We used cross-entropy as the loss function and trained the model using the Adam optimizer [29] with a learning rate that was gradually reduced over the training epochs

TABLE II
EMULATED HIGH-SPEED FROM LOW-SPEED DATA

| Train Data | Test Data | F1 | Note |
|---|---|---|---|
| Low-speed | High-speed | 91 | Zero-shot |
| High-speed | High-speed | 96 | Control-group |

## VI. EXPERIMENTAL VALIDATION

We conducted a series of classification experiments using the dataset described previously. We used the F1 score as the primary evaluation metric, as suggested by Taha et al. [30]. The score is a harmonic mean of precision and recall, calculated as follows:

$$Precision = \frac{TP}{TP+FP} \quad \#(7)$$

$$Recall = \frac{TP}{TP+FN} \quad \#(8)$$

$$F1 = \frac{2*Precision*Recall}{Precision+Recall} \quad \#(9)$$

**Random occlusions** are common in real-world scenarios, such as vehicle scanning, where objects such as street lamps or pedestrians can obstruct the scanning process. We assumed that these occlusions are relatively small in size, and our model needs to be robust to local data loss. Our data sequences include combined scattering angles for each sampling point, which provides our model with the information it needs to locate occluded regions. We used random subsampling, ranging from 5% to 10%, to emulate occlusion scenarios. This procedure increased the size of our dataset by more than a factor of 10. The data was then split into training, validation, and test sets in a 7:3:1 ratio. Our model achieved an F1 score of over 95%. While this score is commendable, it is important to note that our dataset is not very large. A score in this range indicates that the model is capable of performing the task with high fidelity, but higher scores may indicate overfitting

**High-speed sparse data.** As discussed previously, data collected at a turntable speed of 10°/s corresponds to a speed of approximately 10 km/h. A turntable speed of 20°/s corresponds to approximately 20 km/h, which is similar to the speed of slow-moving vehicles. While these speeds may seem moderate, they represent the upper safe limit under the load conditions of our turntable. To explore higher-speed scenarios, we used approximation techniques. High-speed data is very rare, so using low-speed data to train a recognition model can improve data quantity and quality while simplifying data acquisition challenges. To validate our method, we conducted experiments using our acquired and constructed datasets. We referred to data at 10°/s as "low-speed data" and data at 20°/s as "high-speed data". We designed a zero-shot experiment in which a model trained on low-speed data was tested directly on high-speed data. Specifically, we subsampled the low-speed data to emulate high-speed conditions, trained a model on this modified data, and evaluated its performance on the original high-speed data. This is a zero-shot experiment because the model was not exposed to high-speed data during training. The results, shown in Table II, reveal that the model retained some effectiveness even in this challenging scenario.

TABLE III
EMULATED HIGHER-SPEED SCENARIOS

| Equivalent Speed (km/h) | Resampling length | F1 (%) | Note |
|---|---|---|---|
| ≈40 | 30 | 91 | Vehicular cruising speed |
| ≈60 | 20 | 84 | Zero-shot |
| ≈80 | 10 | 71 | Control group |

We then increased the speeds in our emulated scenarios, as shown in Table III, and adjusted the sequence length accordingly. We explored speeds of 40 km/h, 60 km/h, and 80 km/h, which correspond to typical urban vehicle cruising speeds, maximum urban vehicle speeds, and UAV (Unmanned Aerial Vehicles) flight speeds [31], respectively. Remarkably, even under challenging conditions, such as the UAV scenario with only 10 scattering data points per surface, our method maintained an F1 score above 70%. This demonstrates the effectiveness of our method in recognizing surface properties at various speeds, particularly at typical urban cruising speeds.

**Narrow-angle window observations.** We considered scenarios where collecting scattering data across a wide angular range is not feasible. In real-world settings, this may occur when building-obstructed wall surfaces are only accessible through a narrow angular window. It is important to evaluate the performance of our model under these angular constraints. We investigated a 60° angular window, striving to achieve effective recognition with a limited range of angles. We also tried offsetting the angular window relative to the direction of specular reflection (with an offset angle Δ). The results are shown in Table IV. We enforced a constraint of 20 scattering data points within each angle, which aligns with urban vehicle cruising speeds.

It should be noted that the best classification F1 scores were achieved when the angular window enveloped the angles of mirror reflection and was slightly shifted towards the direction of richer features along the normal. Nevertheless, our method consistently achieved F1 scores around 80%, even in the most angularly restrictive scenarios. This experiment demonstrates that our method can maintain high recognition accuracy at low angles, even under medium to low urban scanning speeds, further attesting to the validity of this approach.

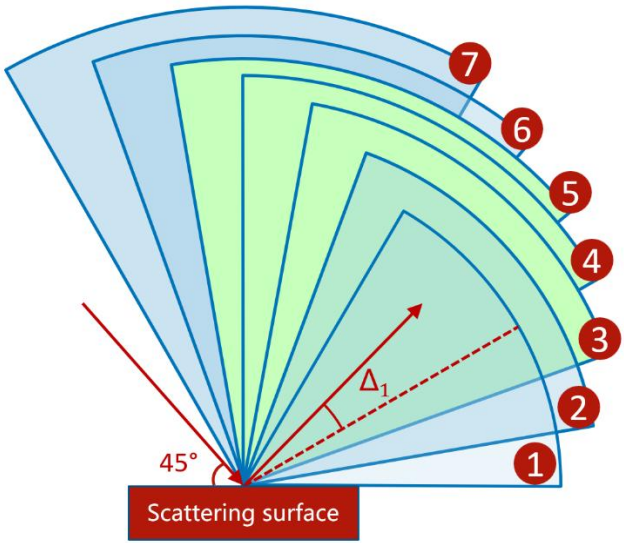

**Fig. 5.** Window positions we attempted under angle-restricted conditions. Each window covers a 60° sector, with two red arrows representing the incident and outgoing waves

TABLE IV
NARROW-ANGLE WINDOW OBSERVATIONS

| Idx. | Offset angle $\Delta$ (°) | Sequence length | F1(%) |
| --- | --- | --- | --- |
| 1 | +15 | 20 | 83 |
| 2 | +5 | 20 | 85 |
| 3 | -5 | 20 | 89 |
| 4 | -15 | 20 | 90 |
| 5 | -25 | 20 | 89 |
| 6 | -35 | 20 | 85 |
| 7 | -45 | 20 | 78 |

## VII. CONCLUSION

In the sphere of integrated sensing and communication, the environmental sensing is becoming as crucial as data transfer. This work proposes an approach in enhancing high-speed scanning capabilities for surface roughness recognition using terahertz frequency. Combining a deep learning framework, this method employs authentic high-speed sampling data to build an automated platform capable of capturing high-fidelity data in scenarios that closely emulate real-world dynamics. The manufacturing and 3D laser characterization of a spectrum of surfaces have facilitated the creation of a rich dataset, important for the training of our specialized neural network. We have aligned our emulated environments with actual laboratory and urban conditions, customizing our model to overcome three typical high-speed challenges: *random occlusion*, *highspeed sparse data*, and *narrow-angle window observations*.

Our experiments have demonstrated the effectiveness of this method, achieving an F1 score of over 95% in scenarios with *random occlusion*. This indicates exceptionally high accuracy within our current dataset, especially in relatively slower situations. Additionally, our approach consistently maintains F1 scores above 90%, 80%, and 70% at emulated vehicular cruising speeds, urban vehicle speed limits, and drone flight speeds, respectively. Notably, we have achieved a 71% F1 score within a 120° range using only 10 randomly positioned scattering data points, highlighting the success of our method in challenging conditions. Moreover, we have identified window positions within a 60° angular range that consistently maintain high speeds, resulting in F1 scores above 90% with 20 randomly positioned scattering data points. The model demonstrates exceptional accuracy, especially in medium to low-speed scenarios and constrained angular windows. Even under extreme conditions, such as very high speeds (e.g. 80 km/h), it maintains a significant degree of reliability. This method may not only prompt the application of terahertz frequency band in surface recognition, but also enriches the broader context of sensing-communication integration, contributing to the evolution of intelligent, high-speed recognition systems.

## APPENDIX A

### SURFACE SAMPLES CHARACTERISTIC

We characterized the surface roughness features based on high-precision point cloud data of the entire surface by the 3D laser scanning technique. Uniform sampling was performed and the calculation range was determined based on the illumination area. The RMS value was calculated based on the surface features within the calculation region. Fig. 6(a) to (g) display the RMS distributions of all seven surfaces. The horizontal and vertical axes represent the offset relative to the central point, and the heatmap values represent RMS values in millimeters (mm). Fig. 6(h) illustrates the effects of all seven surfaces simultaneously in a 3D schematic.

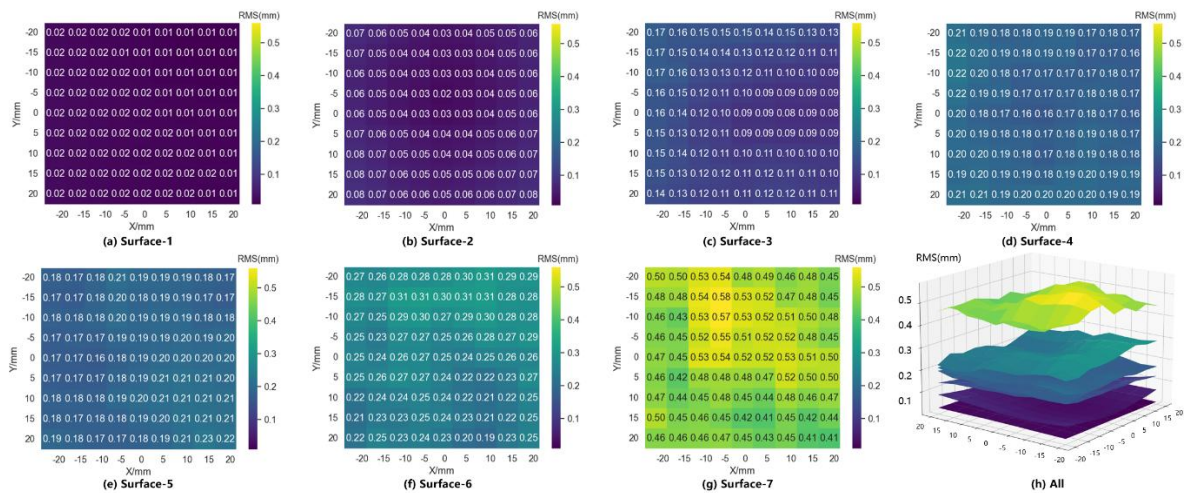

Fig. 6. The heatmaps for RMS values of each surface and the combined display of all seven surfaces in a 3D schematic.